\documentclass[prd,letterpaper,onecolumn,amsmath,amsfonts,amssymb,nofootinbib]{revtex4}
\usepackage {amsmath}
\usepackage[dvips]{graphicx}     
\usepackage{feynmp}
\usepackage{amssymb}
\newcommand{\be}{\begin{equation}}  
\newcommand{\ee}{\end{equation}}
\newcommand{\bear}{\begin{eqnarray}}
\newcommand{\eear}{\end{eqnarray}}

\newcommand{\lapproxeq}{\lower .7ex\hbox{$\;\stackrel{\textstyle
<}{\sim}\;$}}
\newcommand{\gapproxeq}{\lower .7ex\hbox{$\;\stackrel{\textstyle
>}{\sim}\;$}}
\newcommand{\stackdown}[2]{\lower 1.4ex\hbox{$\;\stackrel{\textstyle{#1}}
{\scriptstyle{#2}}\;$}}
\newcommand{\beq}{\begin{equation}}
\newcommand{\eeq}{\end{equation}}

\newcommand{\ba}{\begin{eqnarray}}
\newcommand{\ea}{\end{eqnarray}}

\newcommand{\bea}{\begin{eqnarray}}
\newcommand{\eea}{\end{eqnarray}}

\makeatletter
\def\slash{\@ifnextchar[{\fmsl@sh}{\fmsl@sh[0mu]}}
\def\fmsl@sh[#1]#2{%
  \mathchoice
    {\@fmsl@sh\displaystyle{#1}{#2}}%
    {\@fmsl@sh\textstyle{#1}{#2}}%
    {\@fmsl@sh\scriptstyle{#1}{#2}}%
    {\@fmsl@sh\scriptscriptstyle{#1}{#2}}}
\def\@fmsl@sh#1#2#3{\m@th\ooalign{$\hfil#1\mkern#2/\hfil$\crcr$#1#3$}}
\makeatother
\usepackage{color}

\definecolor{orange}{rgb}{0.9,0.2,0}
\definecolor{brown}{rgb}{0.7,0.3,0.2}
\definecolor{fuxia}{rgb}{1,0,1}
\definecolor{skyblue}{rgb}{0,0.1,0.9}
\definecolor{violetred}{rgb}{0.8,0.13,0.56}
\definecolor{deeppink}{rgb}{1.00,0.08,0.5}
\definecolor{pink}{rgb}{1.00,0.75,0.80}
\definecolor{orchid}{rgb}{0.85,0.44,0.84}
\definecolor{lightpink}{rgb}{1.00,0.71,0.76} 
\definecolor{bluish}{rgb}{0,0.6,0.8}
\begin{document}
\title{INFLATION IN NO-SCALE SUPERGRAVITY }
\author{A. B.~\ Lahanas}
\email{alahanas@phys.uoa.gr}
\affiliation{University of Athens, Physics Department,
Nuclear and Particle Physics Section,
GR--157 71  Athens, Greece}
\author{K. Tamvakis}
\email{tamvakis@uoi.gr}
\affiliation{University of Ioannina,  
Physics Department,
Section of Theoretical Physics, 
GR--451 10  Ioannina, Greece\\
and\\
CERN, Physics Department,
Geneva 23, CH-1211, Switzerland}
\vspace*{2cm}  
\begin{abstract}
$R+R^2$ Supergravity is known to be equivalent to standard Supergravity coupled to two chiral supermultiples with a no-scale K\"ahler potential. Within this framework, that can accomodate vanishing vacuum energy and spontaneous supersymmetry breaking, we consider modifications of the associated superpotential and study the resulting models, which, viewed as generalizations of the Starobinsky model, for a range of the superpotential parameters, describe viable single-field slow-roll inflation. In all models studied in this work the tensor to scalar ratio is found to be small, well below the upper bound established by the very  recent PLANCK and BICEP2  data. 

\end{abstract}
\maketitle
{\bf{Keywords:}} Supergravity, Inflationary Universe, Modified Theories of Gravity

{\bf{PACS:}} 04.65.+e, 98.80.Cq, 04.50.Kd



\section{Introduction }
A phase of de Sitter expansion in the early universe, namely inflation \cite{INFL}, offers an attractive paradigm for the production of primordial curvature perturbations that ultimately generate the large scale structure of the universe \cite{PERT} as well as the resolution of a number of issues of the standard hot big bang cosmology. Although models of inflation typically introduce new scalar fields, it would be intriguing if the extra scalar degrees of freedom were provided by gravity itself. It is known that generalizations of the Einstein Action containing extra powers of the Ricci scalar are equivalent to standard Einstein gravity with an additional scalar field \cite{STELLE}. Specifically the Starobinsky model \cite{STAR}, featuring an extra quadratic $R^2$ term in the Action, is equivalent to standard gravitation with an extra scalar field possessing an inflationary potential \cite{WHITT}. In the point of view of conformal symmetry, expected to hold at the high curvature regime where masses are negligible, an additional motivation to consider the Starobinsky model is supplied by the fact that the quadratic curvature term is bound to be generated by the conformal anomaly at the quantum level. 
Actually, the Starobinsky model is among the few possibilities realizing slow-roll inflation that is in perfect agreement with the PLANCK experiment data \cite{PLANCK}, although its predictions were challenged by  BICEP2 results \cite{BICEP}, claiming the discovery of primordial gravitational waves resulting to a ratio $\, r =0.16^{+0.06}_{-0.05} \,$. In the meantime Planck collaboration released new data of increased precision \cite{PLANCK2}, reconfirming  its previous analyses,  according to which $\, r < 0.11 \,$.   Also BICEP2 and PLANCK  joined collaboration  \cite{Ade:2015tva}  established a robust upper bound $ \, r < 0.12 \, $ which is lower than the value quoted in \cite{BICEP}. 
Independently of the value of $r$, 
it is important to consider and analyze generalizations of it, preferably embedded in a more general framework encompassing the particle physics theories as well. Such a general framework is Supergravity, either as a local extension of supersymmetric particle field theories or as a limit of superstring theory.
Higher derivative supergravity, leading to higher powers of the curvature scalar in the Action, 
has been proven to be equivalent to minimal supergravity coupled to chiral supermultiplets \cite{CECOTTI, KALLOSH}. In particular, the Starobinsky model corresponds to minimal supergravity coupled two chiral supermultiplets. Furthermore, the K\"ahler potential is of the type encountered in {\textit{no-scale models}}  \cite{NOSCALE}. Modifications of the basic no-scale K\"ahler potential and various choices for the superpotential have been studied leading to a number of inflationary alternatives \cite{ELLIS}. All these models are necessarily multifield models containing apart from the inflaton additional scalar fields. These extra fields, coupled to the inflaton through both the potential and their kinetic terms, follow a complicated path in field space towards the minimum. If the inflationary process is to be driven by a single field, these additional scalars should, pressumably, be well settled in their vacuum during inflation.

In this paper we study generalizations of the Starobinsky model derived in the framework of no-scale Supergravity by considering deformations of the basic superpotential. Aiming at the construction of single-field inflationary models that can realize viable slow-roll inflation, we derive a simple correspondence between the desired inflaton potential and the superpotential in a general no-scale Supergravity framework. Such a framework through its geometrical properties allows for spontaneous supersymmetry breaking with a vanishing classical vacuum energy at the Minkowski vacuum. {{Given the no-scale Supergravity framework, we investigate whether viable infationary models can be constructed by considering deformations and modifications of the basic superpotential yielding the Starobisnky model. After reviewing the relation of the no-scale supergravity realization of the Starobinsky model to a scalar field model non-minimally coupled to gravity and the universal attractor, we proceed to consider generalizations that can lead to viable inflationary models. We start with the basic $SU(2,1)/SU(2)\times U(1)$ K\"ahler structure and consider specific one-parameter modified superpotentials, which for a range of the relevant parameter yield viable inflationary behaviour neighboring to that of the Starobinsky model. We also consider models based on a $SU(1,1)/U(1)\times U(1)$ K\"ahler potential and proceed with a particular superpotential, showing that for a certain parameter range, we obtain models with viable slow-roll inflation.}}  

{\section{DERIVATION OF THE STAROBINSKY MODEL FROM NO-SCALE SUGRA}}

The standard no-scale K\"ahler potential involving two chiral superfields $T,\,S$ parametrizing the coset space $SU(2,1)/SU(2)\times U(1)$ is
\be
K\,=\,-3\ln\left(T+\overline{T}\,-|S|^2\,\right)\,.{\label{STANDARDK}}
\ee
For reasons of stability \cite{EKN}, the quadratic $S$-term may be supplemented with extra higher powers of $S$, thus effectively replacing $|S|^2$ in the argument of the logarithm with
 $h(S,\overline{S})\,=\,\overline{S}S\,+\,h_1(\overline{S}S)^2\,+\,\dots$.
The superpotential known to correspond to the Starobinsky potential is
\be
W(S,T)\,=\,W_0\,+\,\lambda\,S\,(T-1)\,,\,{\label{STANDARDW}}
\ee
$\lambda$ being an arbitrary coupling.
The scalar potential is
\be
V\,=\,e^{K}\,\left(\,G_{\overline{i}}\,(K^{-1})^{\overline{i}{j}}\,G_{{j}}\,-3|W|^2\,\right)\,\,\,\,\,{\text{with}}\,\,\,\,\,\,
(K^{-1})^{\overline{i}{j}}\,=\,\frac{e^{-\frac{K}{3}}}{3}\left(\begin{array}{cc}
T+\overline{T}\,&\,S\\
\,&\,\\
\overline{S}\,&\,1
\end{array}\right)
\ee
and
\be\,\,\,\,\,\,G_j\,=\,\frac{\partial W}{\partial\phi^{j}}+W\frac{\partial K}{\partial\phi^{j}}\,=\,\left\{\begin{array}{l}
G_T=\,W_0\,e^{-K/3}\,+\,\lambda\,S\,\left(1\,-3(T-1)e^{-K/3}\right)\\
\,\\
G_S\,=\,\lambda (T-1)\left(1\,+\,3|S|^2e^{-K/3}\right)
\end{array}\right.
\ee
In the limit $S\rightarrow 0$ the Lagrangian reduces to
\be
-3\frac{|\nabla T|^2}{(T+\overline{T})^2}\,-\frac{\lambda^2}{3}\frac{|T-1|^2}{(T+\overline{T})^2}\,
\ee
and can be cast partially in canonical form introducing real scalar fields according to
\be
T\,=\,\frac{1}{2}e^{\sqrt{\frac{2}{3}}\phi}\,+\,i\chi\,.{\label{REAL}}
\ee
Then, it takes the standard Starobinsky form
\be
-\frac{1}{2}(\nabla\phi)^2\,-\frac{\lambda^2}{12}\left(1\,-2e^{-\sqrt{\frac{2}{3}}\phi}\right)^2\,-3e^{-2\sqrt{\frac{2}{3}}\phi}(\nabla\chi)^2\,-\frac{\lambda^2}{3}e^{-2\sqrt{\frac{2}{3}}\phi}\,\chi^2\,,{\label{STARFORM}}
\ee
with the additional presence of the imaginary part field $\chi$. 
Note that $m_{\chi}>|m_{\phi}(\phi)|$ throughout inflation. Further stabilization of the $\chi=0$ vacuum can always be achieved by adding a $(T-\overline{T})^4$ term in the logarithm argument of $K$. 

The fact that supersymmetry is broken through both $G_T\neq 0$ and $G_S\neq 0$ implies that the goldstino corresponds to a mixture of the corresponding supermultiplets. Note also that since at the Minkowski vacuum $G_S$ vanishes, the imposition of a {\textit{quadratic nilpotency constraint \cite{NIL} condition}} $S^2=0$ becomes singular.
\\ \\

{\section{RELATION TO NON-MINIMAL COUPLING AND THE UNIVERSAL ATTRACTOR}}

A scalar field, non-minimally coupled to gravity through a coupling function $\xi\,f(\varphi)\,R$ and having a scalar potential $V(\varphi)$, is described in the Einstein frame with standard Einstein gravity of the matter Lagrangian
\be
-\frac{1}{2}\left(\frac{3}{2}\xi^2\frac{(\nabla f)^2}{(1+\xi f)^2}+\frac{(\nabla\varphi)^2}{1+\xi f}\right)\,-\frac{V}{(1+\xi f)^2}
\ee
In the limit of very large coupling $\xi\,>>\,1$, the second term in the kinetic coefficient is subdominant and the model can be set in the canonical form
\be 
-\frac{1}{2}(\nabla\phi)^2-4\,e^{-2\sqrt{\frac{2}{3}}\phi}\,V(\phi)
\ee
in terms of $\frac{1}{2}e^{\sqrt{\frac{2}{3}}\phi}\,=\,1+\xi\,f$. Then, the model is identical to the Starobinsky model if the scalar potential is
\be 
V(f)\,=\,\frac{\lambda^2}{12}\,\xi^2\,f^2\,.\ee
Therefore, we may conclude that the non-minimal coupling model
\be \frac{1}{2}\left(1+\xi\,f(\varphi)\,\right)\,R\,-\frac{1}{2}\left(\nabla\varphi\right)^2\,-\frac{\lambda^2}{12}\,\xi^2\,f^2(\varphi) {\label{UNI}}
\ee
in the limit of very large coupling $\xi>>1$ is equivalent to the Starobinsky model. In fact, this limit corresponds to a {\textit{universal attractor}} \cite{KALIRO} in which the Starobinsky model predictions for the spectral index and the tensor-to-scalar ratio predictions are indistinguishable from the non-minimal coupling model preditictions, being in agreement with the Planck experiment results. Furthermore, it has also been argued that, taking a non-minimal coupling function $f(\phi)$ different that the function appearing in the potential $V=\lambda^2\,g^2(\phi)$, for sufficiently large $\xi$, we approach again the Starobinsky potential with neighboring inflationary predictions, while for $\phi$ in the inflationary regime the values of the function $g$ approache those of $\xi\,f$ as $g(\phi)\,\approx\,\xi\,f(\phi)\,+\,O(1/\xi)$.
  
The universal attractor model can be embedded in a general no-scale Supergravity model as follows. Consider the K\"ahler potential $K$ and the superpotential $W$
\be
K\,=\,-3\ln\left(2+F(T)+\overline{F}(\overline{T})\,-|S|^2\,\right),\,\,\,\,\,\,\,W\,=\,\lambda\,S\,F(T)\,.{\label{GENNO}}
\ee 
Taking the stabilizer $S\rightarrow\,0$, we obtain the Einstein-frame  Lagrangian
\be 
\frac{1}{2}R\,-3\frac{|\nabla F(T)|^2}{\left(2+F(T)+\overline{F}(\overline{T})\right)^2}\,-\frac{\lambda^2}{3}\frac{|F(T)|^2}{\left(2+F(T)+\overline{F}(\overline{T})\,\right)^2}\,=\,\frac{1}{2}R\,-\frac{3}{4}\left(\frac{\nabla F(\varphi)}{(1+F(\varphi))}\right)^2\,-\frac{\lambda^2}{12}\frac{F^2(\varphi)}{(1+F(\varphi))^2}\,
\ee 
the last expression being valid if we take $F$ to be a real function of $T=\varphi+i\chi$ and set the imaginary part $\chi$ to zero
\footnote{
Throughout this paper when a holomorphic function is defined as "real"  it is meant that  it has real coefficients when it is expanded as a power series. These functions  become  real  when the imaginary part of their  argument  is set to zero .
}
.
This is the Einstein-frame counterpart of the Jordan-frame action  
\be 
\frac{1}{2}\left(1+F(\varphi)\,\right)R\,-\frac{\lambda^2}{12}\,F^2(\varphi)
\ee 
which coincides with ({\ref{UNI}}) in the large non-minimal coupling limit. Thus, in this limit we have a triple correspondence of the general no-scale model defined by ({\ref{GENNO}}), the Starobinsky model and the non-minimal coupling model, all comprising a {\textit{universal attractor}} with coinciding inflationary predictions. 

Nevertheless, since, as we mentioned above, the no-scale model with
$$K=-3\ln(\,2+f(T)+\overline{f}(\overline{T})-|S|^2),\,\,\,\,\,W=\lambda\,S\,g(T)\,,$$
for $g(T)\,\neq\,f(T)$ is expected to give inflationary predictions neighboring to those of the Starobisnsky model in the limit of large $f(T)$, due to the universal attractor properties, a line of generalization would be to start with ({\ref{STANDARDK}}) and {\textit{deform}} the superpotential as
\be 
W\,\propto\,(T-1)^c\,,\,\ee
 the exponent $c$ reducing to the universal attractor case for $c\rightarrow 1$. Further deformations are possible along the lines of 
 \be 
 W\propto\,(T-1)^c(T+b)^d\,.{\label{DEFORM}}
 \ee

\section{GENERALIZATIONS OF THE STAROBINSKY MODEL}

\subsection{$SU(2,1)/SU(2)\times U(1)$}
We consider again the basic $SU(2,1)/SU(2)\times U(1)$ K\"ahler potential ({\ref{STANDARDK}}) expressed in terms of $T$ and $S$ but generalize the superpotential as \cite{NIL}
\be
W\,=\,W_0\,+\,f(T)\,S {\label{GEN-1}}
\ee 
Then, we have
\be 
G_T\,=\,S \frac{\partial f}{\partial T}\,-3e^{-K/3}\left(W_0+  f(T)S\right),\,\,\,\,\,\,\,G_S\,=\,  f(T)\,+\,3e^{-K/3}\overline{S}\left(W_0+ f(T)S\right)
\ee 
or, in the $S\rightarrow 0$ limit,
\be 
G_T=-\frac{3W_0}{(T+\overline{T})},\,\,\,\,G_S=   f(T)\,.
\ee 
In that limit the Lagrangian is
\be 
-\frac{3|\partial T|^2}{(T+\overline{T})^2}\,-\frac{|f(T)|^2}{    
3 (T+\overline{T})^2}\,=\,-\frac{1}{2}(\nabla \phi)^2\,-\,3e^{-2\sqrt{\frac{2}{3}}\phi}(\nabla\chi)^2\,- \frac{1}{3} \,e^{-2\sqrt{\frac{2}{3}}\phi}|f(T)|^2\,,
\ee
where, again, $T=\frac{1}{2}e^{\sqrt{\frac{2}{3}}\phi}+i\chi$.
Again, supersymmetry is broken by both $S$ and $T$ and the Minkowski vacuum corresponds to $f(T_0)=0$. If $f(T)$ is a real function of $T$, we have $|f(T)|^2=|f^*(T^*)|^2=|f(T^*)|^2$ and the scalar potential is an even function of the imaginary part $\chi$, namely, $V(\phi,\,\chi)=V(\phi,\,-\chi)$. This means that the point $\chi=0$ is always a minimum of the scalar potential. We, thus, proceed by setting $\chi=0$ and have

\be 
V(\phi)\,=\, \frac{1}{3} \, e^{-2\sqrt{\frac{2}{3}}\phi}\,f^2(\phi)\,\Longrightarrow\,f(\phi)\,=\, \sqrt{3} \, e^{\sqrt{\frac{2}{3}}\phi}\,\sqrt{V(\phi)}\,.
\ee
This is a useful formula that relates the desired inflationary potential to the function $f(T)$ of the superpotential. In the case of the standard Starobinsky potential ({\ref{STARFORM}}) we obtain
\be
f(\phi)=\frac{\lambda}{2}\left(e^{\sqrt{\frac{2}{3}}\phi}\,-2\right)\,\Longrightarrow\,f(T)\,=\,  \lambda \left(T-1\right)\,
\ee
as expected. 

\begin{figure}[h]
    \centering
    \includegraphics[width=0.5 \textwidth]{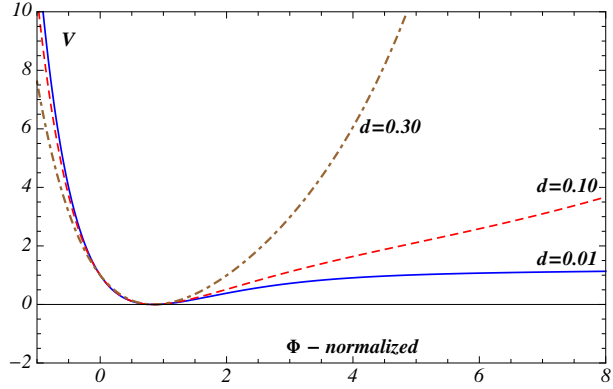} 
    \caption{
    The inflation potential of Eq. (\ref{pot23}), in units of  $\, \lambda^2 / 3 $ ,  for values of the parameters  $c=1, \, b=0 $, and various values of $\, d$. 
        }
    \label{fig1}
\end{figure}

A more general scalar potential of the form
\bea
V(\phi)\,=\,\frac{\lambda^2}{3}e^{-2\sqrt{\frac{2}{3}}\phi}\,\left(\frac{1}{2}e^{\sqrt{\frac{2}{3}}\phi}\,-1\right)^{2c}\left(\frac{1}{2}e^{\sqrt{\frac{2}{3}}\phi}\,+b\right)^{2d}\,{\label{POTENTIAL}}\,,
\label{pot23}
\eea
where $b,\,c,\,d$ are real parameters, corresponds to the superpotential function
\be 
f(T)\,=\,  \lambda \left(T-1\right)^c\left(T+b\right)^d\,,{\label{GENSUPER}}
\ee
being along the lines of the deformation ({\ref{DEFORM}}). The corresponding superpotential is
\be 
W(S,\,T)\,=\,W_0\,+\,{\lambda}\,S\,\left(T-1\right)^c\left(T+b\right)^d\,.{\label{SUPERR}}
\ee
Note that the potential ({\ref{POTENTIAL}}), being 
$$V\,\propto\,\left(1-2e^{-\sqrt{\frac{2}{3}}\phi}\right)^2\,\left(\frac{1}{2}e^{\sqrt{\frac{2}{3}}\phi}\,-1\right)^{2(c-1)}\left(\frac{1}{2}e^{\sqrt{\frac{2}{3}}\phi}\,+b\right)^{2d}$$
possesses an upword tail, in contrast to the completely flat tail of the Starobinsky potential, which could be unstable for superlarge field values. The simpler case of $c=1$ and $b=0$ is shown in Figure \ref{fig1} for various values of the parameter $d$.

The full scalar potential resulting from ({\ref{SUPERR}}), including $\chi=Im(T)$, is
\be 
V(\phi,\,\chi)\,=\,\frac{\lambda^2}{3}e^{-2\sqrt{\frac{2}{3}}\phi}\,\left[\,\left(\frac{1}{2}e^{\sqrt{\frac{2}{3}}\phi}\,-1\right)^{2}\,+\,\chi^2\,\right]^c\,\left[\,\left(\frac{1}{2}e^{\sqrt{\frac{2}{3}}\phi}\,+b\right)^{2}\,+\,\chi^2\,\right]^d{\label{POTENTIAL-1}}\,.
\ee
Single field inflation with this potential assumes a rapid approach to the minimum $\chi=0$. This is the case if the $\phi$-dependent masses $m_{\chi}^2(\phi)\,=\,\frac{1}{6}e^{2\sqrt{\frac{2}{3}}\phi}\frac{\partial^2V}{\partial\chi\partial\chi}$ and $m_{\phi}^2(\phi)\,,=\,\frac{\partial^2V}{\partial\phi\partial\phi}$ are such that $m_{\chi}^2(\phi)>m_{\phi}^2(\phi)$ for values of $\phi$ in the inflationary plateau. In the case $d=0$ the potential posseses an upward tail at superlatge values for $c>1$ reducing to the standard Starobisnsky for $c=1$. The mass ratio and the corresponding inequality is
\be 
\frac{m_{\chi}^2}{m_{\phi}^2}\,=\,\frac{c/2}{\left[(c-1)^2+(3c-4)e^{-\sqrt{\frac{2}{3}}\phi}+4e^{-2\sqrt{\frac{2}{3}}\phi}\right]}\,>\,1\,\Longrightarrow\,\frac{c}{2}\,>\,(c-1)^2\,,{\label{INEQUA}}
\ee
leading to the range $1/2\,\leq \,c\,\leq 2$ in the inflationary regime.

\subsection{$SU(1,1)/U(1)\times U(1)$}
In the seminal paper of Ceccoti \cite{CECOTTI} it was shown that the extension of $N=1$ Supergravity to include a quadratic Ricci curvature term in the Action corresponds to standard Supergravity with two chiral supermultiplets, namely our $T$ and $S$, necessarily introduced through the 
$SU(2,1)/SU(2)\times U(1)$ K\"ahler potential ({\ref{STANDARDK}}).
Another line of generalization of the no-scale models that lead to an inflationary potential in the vicinity of the Starobinsky model is to depart from the full coset space of ({\ref{STANDARDK}}) and restrict ourselves to the $SU(1,1)/U(1)\times U(1)$ structure
\be 
K\,=\,-3\ln(T+\overline{T})\,+\,|S|^2\,.{\label{YOU-ONE}}
\ee
Starting with a general superpotential linear in $S$
\be 
W\,=\,g(T)\,+\,f(T)\,S\,,
\ee
we obtain
$$\begin{array}{l}
G_T\,=\,\frac{\partial g}{\partial T}\,+\,S\,\frac{\partial f}{\partial T}\,-\frac{3}{(T+\overline{T})}\left(\,g+Sf\right)\\
\,\\
G_S\,=\,f\,+\,\overline{S}\left(\,g+Sf\right)
\end{array}
\,\,\Longrightarrow_{S=0}\,
\,\left\{\begin{array}{l}
G_T=\frac{\partial g}{\partial T}\,-\frac{3g}{(T+\overline{T})}\\
\,\\
G_S\,=\,f(T)
\end{array}\right.$$
The limit $S\rightarrow 0$ may be achieved either by imposing a nilpotency condition, if this is feasible, or by introducing an additional stabilizing higher power term in the K\"ahler potential. The resulting scalar potential is
\be 
V(T,\overline{T})\,=\,\frac{1}{3(T+\overline{T})}\left|\frac{\partial g}{\partial T}-\frac{3g(T)}{(T+\overline{T})}\right|^2+\frac{1}{(T+\overline{T})^3}\left(\,|f(T)|^2-3|g(T)|^2\right)\,,{\label{TIPOTE}}
\ee
while, the corresponding kinetic term is
\be 
\frac{3|\nabla T|^2}{(T+\overline{T})^2}\,.
\ee
It is clear that positive-semidefiniteness of ({\ref{TIPOTE}}) is not guaranteed unless $|f(T)|^2-3|g(T)|^2\,=\,c^2\,\geq 0$. Nevertheless, a holomorphic relation between $f$ and $g$ allows only for the choice $c=0$ or
\be 
f(T)\,=\,\sqrt{3}\,g(T)\,\Longrightarrow\,W\,=\,\frac{f(T)}{\sqrt{3}}\,\left(\,1\,+\,\sqrt{3}\,S\,\right)\,.{\label{SUPER-F}}
\ee
Going back to the potential and setting $T=\varphi+i\chi$, we get
\be 
V(\varphi,\,\chi)\,=\,\frac{1}{9(T+\overline{T})}\left|\frac{\partial f}{\partial T}-\frac{3f(T)}{(T+\overline{T})}\right|^2\,=\,\frac{1}{18\varphi}\left|\frac{\partial}{\partial\varphi} f(\varphi+i\chi)\,-\frac{3}{2\varphi}f(\varphi+i\chi)\,\right|^2\,.
\ee
Assuming that $f(T)$ {\textit{is a real function of $T$}}, it is clear that the potential will be an even function of the imaginary part $\chi$, i.e.
\be 
V(\varphi,\,\chi)\,=\,V(\varphi,\,-\chi)\,.\ee
This implies that the point $\chi=0$ will be always a minimum. Therefore, we can proceed setting $\chi=0$. Then, the potential takes the form
$$V(\varphi)\,=\,\frac{1}{18\varphi}\left(f'(\varphi)-\frac{3}{2\varphi}f\right)^2\,=\,\frac{1}{3}\left(\,\varphi^{3/2}\left(\varphi^{-3/2}f\right)'\,\right)^2\,=\,\frac{\varphi^2}{18}\left(\,\left(\varphi^{-3/2}f(\varphi)\right)'\,\right)^2$$
or
\be 
f(\varphi)\,=\,\sqrt{18}\,\varphi^{3/2}\,\int\,\frac{d\varphi}{\varphi}\,\sqrt{V(\varphi)}\,.{\label{F-1}}
\ee
The field $\varphi=Re(T)$ can be replaced by the canonical field {{$\phi\,=\,\sqrt{\frac{3}{2}}\,\ln 2\varphi$ or $\varphi\,=\,\frac{1}{2}e^{\sqrt{\frac{2}{3}}\phi}$}}, in terms of which $f(\phi)$ reads
\be 
{{f(\phi)\,=\,\frac{\sqrt{6}}{2}\,e^{\sqrt{\frac{3}{2}}\phi}\,\int\,d\phi\,\sqrt{V(\phi)}\,.}}{\label{F-2}}
\ee 
The formulae ({\ref{F-1}}) and ({\ref{F-2}}) are usefull expressions from which we may infer by analytic continuation the superpotential function $f(T)$. Note that supersymmetry is broken through $G_S(T)=f(T)\neq 0$ and $W\neq 0$, while $G_T(T)$ vanishes at the Minkowski vacuum. $S$ can be identified with the goldstino multiplet and the limit $s=0$ could be obtained through a nilpotency condition $S^2=0$. Alternatively stabilization can always be achieved by including quartic $S$-terms in the K\"ahler potential.

As a first application of the above we may consider the standard Starobinsky scalar potential
$$V(\varphi)\,=\,\frac{\lambda^2}{12}\left(1-\frac{1}{\varphi}\right)^2$$
and obtain 
\be 
f(T)\,=\,\lambda\sqrt{\frac{3}{2}}\sqrt{T}\left(\,T\ln T\,+1\,+\,C\,T\,\right)\,\Longrightarrow\,W\,=\,\frac{\lambda}{\sqrt{2}}\sqrt{T}\left(\,T\ln T\,+1\,+\,C\,T\,\right)\,\left(\,1\,+\,\sqrt{3}\,S\,\right)\,.\ee
As another example we consider the potential
\be V(\varphi)\,=\,\frac{\lambda^2}{12}\left(\,1\,-\frac{1}{\varphi}\right)^2\,\varphi^{2d}\,.{\label{TAIL}}
\ee
Applying ({\ref{F-1}}) for ({\ref{TAIL}}) we obtain ($C$ is an integration constant)
\be 
f(T)\,=\,\frac{\lambda\sqrt{3}}{d\sqrt{2}}\,T^{\frac{1}{2}}\,\left(\,C\,d\,T\,+\,T^{1+d}\,+\,\frac{d}{1-d}T^d\right)\,.{\label{SUPER-D}}
\ee

The realization of single-field inflation depends crucially on how fast the vacuum value of the imaginary part $\chi=0$ is approached during the inflationary phase. The field dependent mass of $\chi$ is 
\be 
m_{\chi}^2\,=\,\frac{2}{3}\varphi^2\,\frac{\partial^2V}{\partial\chi\partial\chi}\,=\,\frac{2\varphi}{27}\left(\,\left(f''(\varphi)-\frac{3}{2\varphi}f'(\varphi)\right)^2\,-\left(f'''(\varphi)-\frac{3}{2\varphi}f''(\varphi)\right)\left(f'(\varphi)-\frac{3}{2\varphi}f(\varphi)\right)\,\right)\,.
\ee
The mass of the inflaton is more conveniently expressed in terms of the canonical field $\phi$ as
\be 
m_{\phi}^2\,=\,\frac{\partial^2V(\phi)}{\partial\phi\partial\phi}\,=\,\frac{\partial^2}{\partial\phi^2}\left(\,{{\frac{2}{3}}}\left(\,\left(\,f\,e^{-\sqrt{\frac{3}{2}}\phi}\right)'\,\right)^2\,\right)\,,
\ee
where all derivatives are with respect to the canonical field. 
For the last example above, for large $\phi$, we have $f(\varphi)\sim\frac{\lambda\sqrt{3}}{d\sqrt{2}}\varphi^{3/2+d}$ and
$$\begin{array}{l}
m_{\phi}^2\,\approx\,\frac{2d^2\lambda^2}{9}\,\varphi^{2d}\\
\,\\
m_{\chi}^2\,\approx\,\frac{\lambda^2}{9 d^2}\left(d+\frac{3}{2}\right)\left(\,\left(d-\frac{1}{2}\right)^2\,+\,\frac{5}{4}\right)\,\varphi^{2d}
\end{array}$$
which necessitates the inequality
\be \left(d+\frac{3}{2}\right)\left(\,\left(d-\frac{1}{2}\right)^2\,+\,\frac{5}{4}\right)\,>\,2d^4\,{\label{RANGE-D}}
\ee
for the  $\, m_{\chi}^2 > m_{\phi}^2 $ to hold.

\section{DUAL DESCRIPTION}  
The Starobinsky model ({\ref{STARFORM}}) for $\chi=0$ can also be set in its original Ricci curvature form by going back to the Jordan frame
\be 
\frac{1}{4}R\,e^{\sqrt{\frac{2}{3}}\phi}\,-\frac{\lambda^2}{12}
\left(\frac{1}{2}e^{\sqrt{\frac{2}{3}}\phi}\,-1\right)^2\,=\,\frac{1}{2}R\,+\,\frac{\alpha}{2}R^2\,,{\label{STARFORM-1}}
\ee
with $\alpha=3/2\lambda^2$. An alternative line of generalization leading to a more general potential like ({\ref{POTENTIAL}}) can arise from a generalization of ({\ref{STARFORM-1}}). We may start with the Action
\be 
{\cal{S}}\,=\,\int\,d^4x\,\sqrt{-g}\,\left\{\,\frac{1}{2}R\,+\,\frac{\alpha}{2}R^{2(1-n)}\,\right\}{\label{GENR}}\,.
\ee
This can also be written as
\be
{\cal{S}}_J\,=\,\int\,d^4x\,\sqrt{-g}\,\left\{\,\frac{1}{2}R\left(1+2\Phi\right)\,-C\,\Phi^{\frac{2(1-n)}{(1-2n)}}\,\right\}{\label{JORDAN}}
\ee
with 
\be
C\,=\,\frac{1}{2\alpha^{1/(1-2n)}}\frac{(1-2n)}{(1-n)^{\frac{2(1-n)}{(1-2n)}}}\,\,\,\,\,\,\,.
\ee
The Einstein-frame Action corresponding to ({\ref{JORDAN}}) is
\be
{\cal{S}}_E\,=\,\int\,d^4x\,\sqrt{-g}\,\left\{\,\frac{1}{2}R\,-\frac{3}{4}\left(\nabla\ln(1+2\Phi)\right)^2\,-C\frac{\Phi^{\frac{2(1-n)}{(1-2n)}}}{(1+2\Phi)^2}\,\right\}
\ee
or, introducing again the canonical field $\phi$ as
\be
1+2\Phi\,=\,\frac{1}{2}e^{\sqrt{\frac{2}{3}}\phi}\,,{\label{FI}}
\ee
\be 
{\cal{S}}_E\,=\,\int\,d^4x\,\sqrt{-g}\,\left\{\,\frac{1}{2}R\,-\frac{1}{2}(\nabla\phi)^2\,-\frac{C}{2^{\frac{(1-n)}{1-2n}}}\left(1\,-2e^{-\sqrt{\frac{2}{3}}\phi}\right)^2\,\left(\,\frac{1}{2}e^{\sqrt{\frac{2}{3}}\phi}\,-1\right)^{\frac{2n}{(1-2n)}}\,\right\}
\ee
This corresponds to case
\be
d=0,\,\,\,c\,=\,\frac{1-n}{1-2n},\,\,\,\,\frac{\lambda^2}{12}=\frac{(1-2n)}{2\alpha^{1/(1-2n)}(2(1-n)^2)^{\frac{(1-n)}{(1-2n)}}}
\ee
of ({\ref{POTENTIAL}}). The parameter $n$ can be positive $0\leq n\leq 1/2$, giving an exponent of $R$ between $1$ and $2$, or negative giving an exponent $2(1+|n|)\geq 2$. Note however that in the second case the potential vanishes asymptotically. This model is embeddable in a no-scale supergravity model with superpotential
\be
W\,=\,W_0\,+\,\frac{\lambda}{\sqrt{3}}\,S\,\left(T-1\right)^{\frac{(1-n)}{(1-2n)}}\,.
\ee

\section{SLOW-ROLL INFLATION}

Based on the generalizations of the Starobinsky model discuss in the previous sections we may proceed now and study their inflationary predictions. For a canonical field in an FRW background, the equations of motion are 
\be 
H^2\,=\,\frac{8\pi}{3M^2}\rho,\,\,\,\,\,\,\,\,\,\,\,\,\ddot{\phi}+3H\dot{\varphi}+V'(\phi)=0\,.\ee
The parameters relevant to single-field inflation are defined as\footnote{Note that the definition for $\eta$ differs from the {\textit{potential definition}} $\eta_V\,=\,\frac{M^2}{8\pi}\left(\frac{V''}{V}\right)$
as $\eta\,=\,\eta_V\,-\epsilon\,.$}
\be \epsilon\,=\,\frac{3}{2}\left(\frac{p}{\rho}+1\right)\,=\,\frac{4\pi}{M^2}\left(\frac{\dot{\phi}}{H}\right)^2\,=\,-\frac{\dot{H}}{H^2},\,\,\,\,\,\,\,\,\,\,
\eta\,=\,-\frac{\ddot{\phi}}{H\dot{\phi}}\,.
\ee
Assuming that the {\textit{slow-roll approximation}} holds, we may have
$$\dot{\phi}^2<<V\,\Longrightarrow\,H^2\,\approx\,\frac{8\pi}{3M^2}V,\,\,\,\,\,\,\dot{\phi}\,\approx\,-\frac{V'(\phi)}{3H}\,.$$
Then, we have the following approximate expressions for the {\textit{slow-roll parameters}}
\be 
\epsilon\,\approx\,\frac{M^2}{16\pi}\left(\frac{V'(\phi)}{V(\phi)}\right)^2,\,\,\,\,\,\,\,\, 
\eta\,\approx\,\frac{M^2}{8\pi}\left(\frac{V''(\phi)}{V(\phi)}\,-\frac{1}{2}\left(\frac{V'(\phi)}{V(\phi)}\right)^2\,\right)\,.
\ee
\subsection{A $S\bf{U(1,1)/U(1)\times U(1)}$ Example}
Let's consider the potential 
\bea 
V(\phi)\,=\,\frac{\lambda^2}{12}\left(1-2e^{-\sqrt{\frac{2}{3}}\phi}\right)^2\,e^{2d\sqrt{\frac{2}{3}}\phi}\,=\,\frac{\lambda^2}{12}2^{2d}\left(1-\frac{1}{\varphi}\right)^2\,\varphi^{2d}\,,
\label{pot54}
\eea
written in terms of the canonical field $\phi$ or $\varphi=\frac{1}{2}e^{\sqrt{\frac{2}{3}}\phi}$. This potential arises in the case of the K\"ahler metric ({\ref{YOU-ONE}}) and corresponds to a superpotential
$$W\,\propto\,(1+\sqrt{3}S)\,T^{1/2}\left(dCT+T^{1+d}+\frac{d}{1-d}T^d\right)\,.$$
We obtain
\be \begin{array}{l}
\epsilon\,\approx\,\frac{M^2}{6\pi}\left(\,d\,+\,\frac{2e^{-\sqrt{\frac{2}{3}}\phi}}{1-2e^{-\sqrt{\frac{2}{3}}\phi}}\,\right)^2\,=\,\frac{M^2}{6\pi}\left(d+\frac{1}{\varphi-1}\right)^2\,\\
\,\\ 
\eta\,\approx\,\frac{M^2}{6\pi}\left(\,d^2\,+\,\frac{2(2d-1)e^{-\sqrt{\frac{2}{3}}\phi}}{1-2e^{-\sqrt{\frac{2}{3}}\phi}}\,\right)\,=\,\frac{M^2}{6\pi}\left(\,d^2\,+\,\frac{(2d-1)}{\varphi-1}\,\right)\,
\end{array}
\ee

\begin{figure}[h]
    \centering
    \includegraphics[width=0.6 \textwidth]{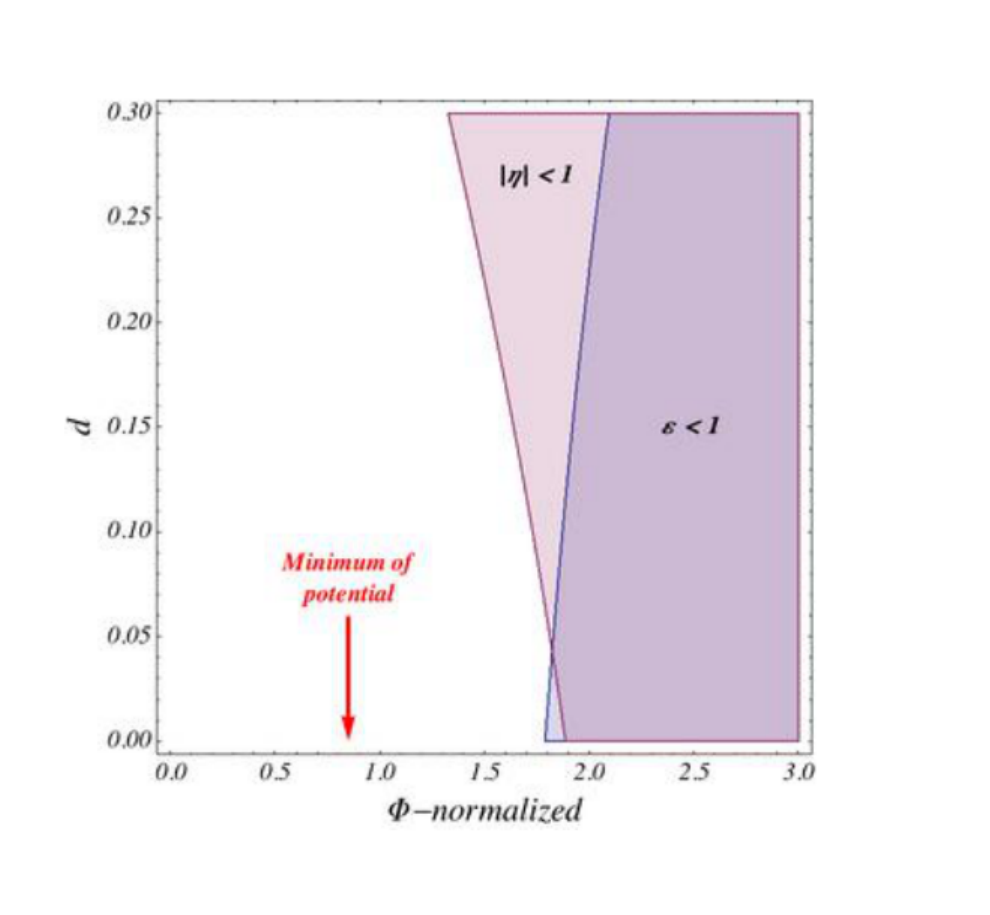} 
    \caption{
    Within the shaded regions in the  $\phi, d   $  plane,   for the potential given by Eq. (\ref{pot54}) ,   the slow-roll parameters 
    are $\, | \eta | < 1 \,$    and  $\, \epsilon  < 1 \,$ and slow-roll holds.  The value of the field minimizing the potential is also shown. 
        }
    \label{fig2}
\end{figure}

\begin{figure}[h]
    \centering
    \includegraphics[width=0.45 \textwidth]{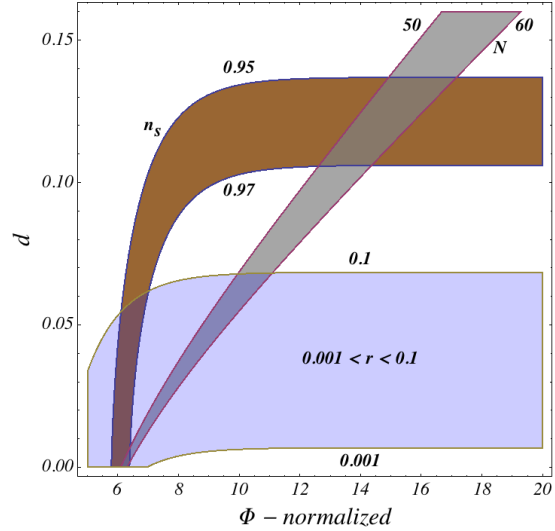} 
    \caption{
    In the $\phi, d   $  plane,  and for the potential given by Eq. (\ref{pot54}),  we delineate the regions ( shaded ) for
    which $ 50 < N < 60 $, $\,  0.95 < n_s < 0.97$ and $\, 0.001 < r < 0.1$. 
    Only for small values of $\, d$ the three quantities can be simultaneously within observational limits.  
        }
    \label{fig3}
\end{figure}

The amount of required inflation is parametrized in terms of the {\textit{number of e-folds}} given by the integral formula
\be N\,=\,-\int_{\phi}^{\phi_1}d\phi\frac{V}{V'} {\label{EFOLDS}}\ee
in terms of the initial field value $\phi$ and the field value at the end of inflation $\phi_1$, defined by the breakdown of the slow-roll approximation ($\epsilon(\phi_1)\,\geq\,1$). Integrating the expression ({\ref{EFOLDS}}) we obtain
\be 
\begin{array}{l}
N\,=\,\frac{3}{4(1-d)}\ln(\varphi_1/\varphi)\,-\frac{3}{4d(1-d)}\ln\left(\frac{\varphi_1-1+\frac{1}{d}}{\varphi-1+\frac{1}{d}}\right)\\
\,\\
\,=\,\frac{1}{2{{(1-d)}}}\sqrt{\frac{3}{2}}(\phi_1-\phi)\,-\frac{3}{4d(1-d)}\ln\left(\frac{e^{\sqrt{\frac{2}{3}}\phi_1}+2\left(\frac{1}{d}-1\right)}{e^{\sqrt{\frac{2}{3}}\phi}+2\left(\frac{1}{d}-1\right)}\right)
\end{array}
\ee
The field value $\phi_1$ is determined by
$$\frac{4}{3}\left(d+\frac{1}{\varphi_1-1}\right)^2\, =  \,1\,\,\Longrightarrow\,\varphi_1 = \,\frac{1+\frac{2}{\sqrt{3}}(1-d)}{1{{-}}\frac{2d}{\sqrt{3}}}\,\,\,\,\,\,{\text{or}}\,\,\,\,\,\phi_1\, = \,\sqrt{\frac{3}{2}}\,\ln\left(\frac{2+\frac{4}{\sqrt{3}}(1-d)}{1{{-}}\frac{2d}{\sqrt{3}}}\right)\,$$
with $\phi_1$ larger than the field value at the minimum $\phi_0$, obtained from $2e^{-\sqrt{\frac{2}{3}}\phi_0}=1$ or $\varphi_0=1$
\footnote{
This actually holds for values  $d < 1/2 $ which is the regime in which agreement with obsercvational data can be obrtained. 
The region for which both $\epsilon$ and $\eta$ are less than unity is shown in figure \ref{fig2}. One observes that $\epsilon < 1$ results to $\, | \eta | < 1 $, as well, for values  $ \, d > \frac{ 1+\sqrt{3}-\sqrt{7} }{2} \approx 0.043$. For smaller values of $d$  departure from slow-roll  is determined by $\, | \eta(\varphi_1) | = 1 $. In this case the resulting value $\varphi_1$  is very close to the one determined above as shown in the same figure.  
}.
In Figure \ref{fig2}  we display the regions of the field valued  slow-roll parameters, when $d < 0.3$,  for which slow-roll approximation is valid. In Figure \ref{fig3}, for values of $ d < 0.15$, we show the regions (shaded)  for  the number of e-foldings left,  $\,N$, the  scalar tilt index $\, n_s$ and the tensor to scalar ratio $\, r$,    

\be  
n_s\,=\,1\,-4\epsilon\,+\,2\eta,\,\,\,\,\,\,r\,=\,16\,\epsilon\,.   
\ee

The acceptable value for  $\, N$,  should be in the range $\, 50 - 60 $.  As for the spectral index,  Planck data combined with WMAP large-angle polarization  yield $\,  n_s = 0.9603 \pm 0.0073 \, $, which is robust to the addition of external data sets. 
This value is only slightly changed, in view of the new Planck data \cite{PLANCK2},  $\,  n_s = 0.968 \pm 0.006 \, $.  Thus 
the most  favourable value for the spectral index is very close  to $\, n_s \approx 0.96 $.  
For the much disputed tensor to scalar ratio $\, r$ of the primordial spectrum,  PLANCK collaboration and BICEP2 have now established a  joined  upper bound,  $\, r < 0.12 \, $  \cite{Ade:2015tva}, and thus values of  $r$ lower than or close to  $ \,  \sim 0.1$ are consistent with current observations. 
In Figure \ref{fig3} we display regions (as shadowed) in which $\, n_s$ is  in the range $\, 0.95 - 0.97  $ and the ratio $\, r$ is less than $0.1$ and larger than $0.001$.  The region where $\,  50 < N < 60 \, $ is also shown. The latter overlaps with the $\, n_s$ region at two 
distinct areas specified by values of the parameter $\, d \sim 0.12$ and $\, d \lesssim 0.01$.  Within the first region $\, r \sim 0.25$ and in the second $\, r \lesssim 0.006$.  In this sense we may say that the particular  model interpolates between values favoured by chaotic inflation scenarios, which predict $\, r \sim 0.1$, and Starobinsky's model predictions that yield small values, $\, r \sim 0.001$, for the ratio $r$. Within the area shown in Figure \ref{fig3}  the slow-roll parameters  are well within the slow-roll regime  $\, | \eta | << 1 \,$    and  $\, \epsilon  << 1 \,$.  Note that the bound  $ r < 0.1 $  excludes the overlapping region for $n_s, \,  N$,  centered around values  of $d$ in the vicinity of $\, d \sim 0.12$,  leaving as the only possibility  the region designated by small values $ \, d < 0.01$.

\subsection{A $\bf{SU(2,1)/SU(2)\times U(1)}$ Example}

Consider now the potential
\bea
V\,=\,\frac{\lambda^2}{3}\,e^{-2\sqrt{\frac{2}{3}}\phi}\,\left(\,\frac{1}{2}e^{\sqrt{\frac{2}{3}}\phi}\,-1\right)^{2c}
\label{modelb}
\eea
arising in the case of the $SU(2,1)/SU(2)\times U(1)$ K\"ahler metric with a superpotential
\be 
W\,=\,W_0\,+\,\lambda S(T-1)^c\,.{\label{SUPER-C}}
\ee 
Note that for $c=1$ we have exactly the Starobisky potential. Note also that this model corresponds to the $R+R^m$ theory with exponent $m=2c/(2c-1)$. We have
$$\frac{V'}{V}\,=\,2\sqrt{\frac{2}{3}}\left(\,-1+\frac{\frac{c}{2}e^{\sqrt{\frac{2}{3}}\phi}}{\frac{1}{2}e^{\sqrt{\frac{2}{3}}\phi}-1}\right)\,=\,2\sqrt{\frac{2}{3}}\left(-1+\frac{c\varphi}{\varphi-1}\right)$$
and
\be 
\epsilon(\phi)\,=\,\frac{4}{3}\left(\,-1+\frac{\frac{c}{2}e^{\sqrt{\frac{2}{3}}\phi}}{\frac{1}{2}e^{\sqrt{\frac{2}{3}}\phi}-1}\right)^2\,=\,\frac{4}{3}\left(-1+\frac{c\varphi}{\varphi-1}\right)^2,\,\,\,\,\,\,\,\eta\,=\,\frac{4}{3}\frac{\left[\left((c-1)\varphi+1\right)^2-c\varphi\right]}{\left(\varphi-1\right)^2}
\ee
\begin{figure}[h]
    \centering
    \includegraphics[width=0.5 \textwidth]{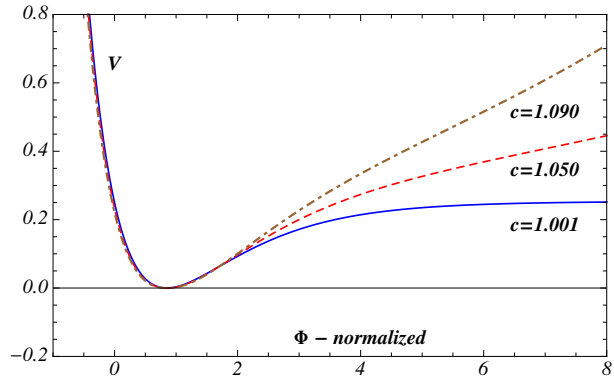} 
    \caption{
    As in Figure \ref{fig1} for the 
    inflation potential of Eq. (\ref{modelb}) for values of the parameters  $c= 1.001, 1.050, 1.090 $.. 
        }
    \label{fig11}
\end{figure} 
The end of inflation is given by $\epsilon(\varphi_1)=1$ and it occurs at a field value
$$\varphi_1\,=\,\frac{1}{1-\frac{c}{\left(1+\frac{\sqrt{3}}{2}\right)}}\,.$$
The number of e-folds is given by
\be 
N\,=\,\frac{3}{4}\ln\varphi_1-\frac{3}{4}\frac{c}{(c-1)}\ln\left((c-1)\varphi_1+1\right)\,-\frac{3}{4}\ln\varphi\,+\,\frac{3}{4}\frac{c}{(c-1)}\ln\left((c-1)\varphi+1\right)\,.
\ee
Taking sample values for the exponent $c$ we may arrive at the values for the inflationary parameters shown in Table \ref{tabit}. 

\begin{table} 
\begin{center}
\begin{tabular}{|l|l|l|l|l|l|l|l|}
\hline
$c$&$\phi_1$&$N$&$\phi$&$\epsilon$&$\eta$&$n_s$&$r$\\
\hline
$1.001$&$1.79$&$59.87$&$6.29$&$0.00022$&$-0.01585$&$0.96740$&$0.00356$\\
\hline
$1.01$&$1.80$&$55.01$&$6.71$&$0.00046$&$-0.01098$&$0.97622$&$0.00730$\\
\hline
$1.05$&$1.86$&$55.67$&$9.31$&$0.00347$&$+ 0.00207$&$0.99025$&$0.05560$\\
\hline
$1.08$&$1.91$&$59.68$&$12.13$&$0.00856$&$+0.00841$&$0.98260$&$0.13690$\\
\hline
$1.09$&$1.92$&$59.55$&$12.98$&$0.01081$&$+0.01074$&$0.97823$&$0.17301$\\
\hline
\end{tabular}
\caption{Sample values for the inflationary parameters, for various values of the parameter $\, c$, for the potential given by Eq. (\ref{modelb}).}
\label{tabit}
\end{center}
\end{table} 

\begin{figure}[h] 
    \centering
    \includegraphics[width=0.57 \textwidth]{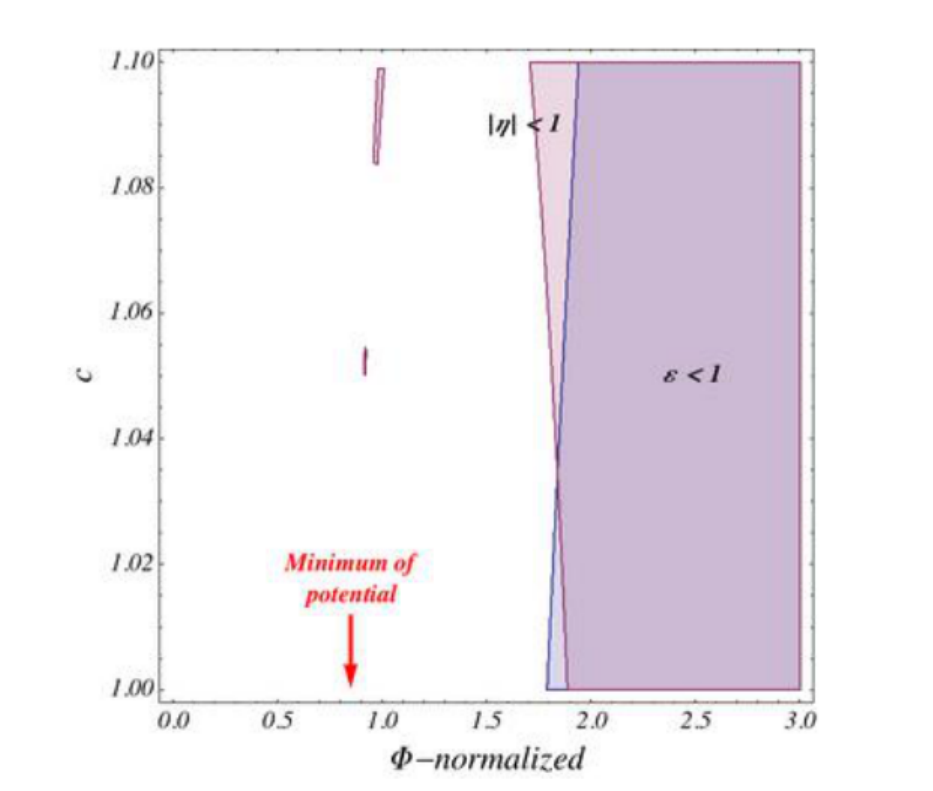} 
    \caption{
    As in Figure \ref{fig2} for the potential of Eq. (\ref{modelb}). 
    Within the shaded regions,  in the  $\phi, c   $  plane,   the slow-roll parameters 
    are $\, | \eta | < 1 \,$    and  $\, \epsilon  < 1 \,$ and slow-roll holds.  The value of the field minimizing the potential is also shown. 
        }
    \label{fig22}
\end{figure}

\begin{figure}[h]
    \centering
    \includegraphics[width=0.45 \textwidth]{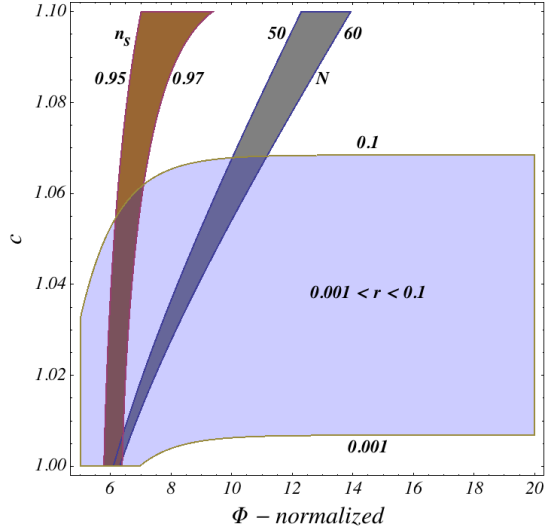} 
    \caption{
     As in Figure \ref{fig3},  for the potential of Eq. (\ref{modelb}), 
    we delineate regions ( shaded )  in the $\phi, c   $  plane   for
    which $ 50 < N < 60 $, $\,  0.95 < n_s < 0.97$ and $\, 0.001 < r < 0.1$. 
    Only for values of $\, c$ in the vicinity of $1$ the three quantities can be simultaneously within observational limits.  
        }
    \label{fig33}
\end{figure}

In Figure \ref{fig11} we plot the shape of the potential for this model, given by Eq. (\ref{modelb}), for three representative value of the parameter $\, c$.  In Figure \ref{fig22} we display the regions (as shaded ) where the slow-roll parameters are less than unity and slow-roll holds. In Figure \ref{fig33}, in the $\, \phi, c$ plane,  we display the regions for the number of foldings $\, N$ and the parameters $n_s$ and $\,r$. As in the case of the potential (\ref{pot54}),  considered previously,  the shadowed regions correspond to $\, 50 < N <  60$, 
$0.95 < n_s < 0.97$ and $\, 0.001 < r < 0.1$.  The region in which the spectral index and the number of e-foldings is within observational limits forces $\, c$ to values $\, c < 1.01 $ and therefore the potential (\ref{modelb})  deviates little from the Starobinsky 's model .  The maximum value of $\, r$ within the allowed region in this case is $\, r \simeq 0.006 $. 



\section{BRIEF SUMMARY AND CONCLUSIONS}
{{The obvious need to embed inflationary models into the general framework of fundamental particle physics leads naturally to the consideration of these models in the framework of Supergravity. In the light of existing CMB data, favouring large field inflation, among the possible viable inflationary models the Starobinsky model has received particular attention, realizing the attractive property that the inflaton degree of freedom is supplied by gravity itself. The minimal Supergravity theory incorporating quadratic curvature terms has been shown to be equivalent to standard minimal Supergravity coupled to a pair of chiral multiplets \cite{CECOTTI, KALLOSH}. Furthermore, the associated K\"ahler potential is that of the $SU(2,1)/SU(2)\times U(1)$ {\textit{no-scale models}} \cite{NOSCALE}. The geometrical properties of these models can accomodate a naturally vanishing classical vacuum energy and spontaneous supersymmetry breaking. For the given K\"ahler potential, the Starobinsky model is obtained for a particular choice of superpotential. Nevertheless, it is legitimate to investigate whether viable inflationary models can arise within this general context from various deformations or modifications of the superpotential.}}

In the present article, after briefly reviewing the relation of the no-scale supergravity realization of the Starobinsky model to a scalar field model non-minimally coupled to gravity and the universal attractor, we proceeded to consider generalizations that can lead to viable inflationary models. We started with the basic $SU(2,1)/SU(2)\times U(1)$ K\"ahler structure and considered superpotentials 
of the form (25) that lead to generalizations of the Starobisnky potential having the generic form (26).  In those models 
the imaginary part of the chiral superfield $T$ settles to its vacuum value at the origin and the scalar potential reduces to an 
effectively  single field potential, the inflaton being the real part of $T$.  This was exemplified analytically in superpotentials (61)  by showing that in the parameter range $1/2 < c < 2$, the imaginary part of the chiral superfield $T$ has a field dependent  mass which is large enough, as compared with that of the inflaton,  to drive it towards its minimum value at $\, Im \, T =0$, reducing the potential to (\ref{modelb}).  Interestingly enough the Starobinsky potential is recovered for the value  $\, c=1$,  being within the aforementioned range of the values of $\, c$.  We then  proceeded to study slow-roll inflation of this model and found that the parameter $c$  must be quite  close to the Starobinsky value $c=1$ in order to yield inflationary predictions as shown in Figure \ref{fig33}. Within this region the tensor to scalar ratio $\, r$ cannot exceed $\, 0.006$.  

We also considered modifications of the basic K\"ahler structure to $SU(1,1)/U(1)\times U(1)$, assuming a general superpotential of the form ({\ref{SUPER-F}}). We proceeded to consider a particular form  ({\ref{SUPER-D}}) , characterized by one parameter $d$ that sets the scale of  departure from the inflationary plateau,  and showed that for the parameter range defined by ({\ref{RANGE-D}}) the scalar potential reduces to ({\ref{pot54}}). We proceeded to study slow-roll inflation of this model. The results can be exhibited in Figure \ref{fig3}.  In this model the number of e-foldings and the spectral index can be within observational limits for values of the  
parameter $\, d$ around $\, \sim 0.12$ and also  small values of it,  $\, d < 0.01$. In the "high"  $d$ regime ( $d \sim 0.12$ ) the values of $\,r$ are  large $\, r > 0.2 \, $ and thus the predictions mimic those of the chaotic inflation models.  Such large values however  are above the bounds set by the recent  PLANCK and BICEP2 data.  The most reasonable option for the range of $d$ is the "low" region, $ \, d < 0.01$,  which inevitably leads to predictions close to the Starobinsky model with values of $r$  that can be slightly enhanced, reaching   $\, r \simeq 0.006$. 
      
\vspace*{5mm}  
{\textbf{Acknowledgements}}   

This research has been co-financed by the European Union (European Social Fund - ESF)
and Greek national funds through the Operational Program Education and Lifelong
Learning of the National Strategic Reference Framework (NSRF) - Research Funding
Program: {\textit{THALES-Investing in the society of knowledge through the European Social Fund.}}  K.T. would also like to thank CERN Theory Division for the hospitality and I. Antoniadis and A. Kehagias for discussions.

\end{document}